\documentclass[a4paper,fleqn]{latex_styles/cas-sc}
\usepackage[authoryear,longnamesfirst]{natbib}

\usepackage{subfigure}
\usepackage{letltxmacro}
\usepackage{hyperref}
\usepackage{amsmath}
\usepackage{booktabs}
\usepackage{longtable}
\usepackage{array}
\usepackage{multirow}
\usepackage{wrapfig}
\usepackage{colortbl}
\usepackage{pdflscape}
\usepackage{tabu}
\usepackage[normalem]{ulem}
\usepackage{makecell}
\usepackage{xcolor}
\usepackage{caption}
\usepackage{tabularx}
\usepackage{adjustbox}
\usepackage{natbib}
\usepackage{mfirstuc}
\usepackage{float}
\usepackage{placeins}
\usepackage{footnote}
\bibpunct{(}{)}{,}{a}{}{;}

\DeclareCaptionFormat{bold}{\textbf{#1#2}#3}
\setcitestyle{aysep={}}
\setcitestyle{authoryear,open={(},close={)}}

\def\tsc#1{\csdef{#1}{\textsc{\lowercase{#1}}\xspace}}
\tsc{WGM}
\tsc{QE}
\tsc{EP}
\tsc{PMS}
\tsc{BEC}
\tsc{DE}
\usepackage{siunitx}

\begin{document}

\let\WriteBookmarks\relax
\def\floatpagepagefraction{1}
\def\textpagefraction{.001}

\shorttitle{\textbf{Consumption and capital growth}}

\shortauthors{Gordon Getty}

\title [mode = title]{\textbf{Consumption and capital growth}}                      
% \tnotemark[1,2]

% \tnotetext[1]{This document is the results of the research
%   project funded by the National Science Foundation.}

% \tnotetext[2]{The second title footnote which is a longer text matter
%   to fill through the whole text width and overflow into
%   another line in the footnotes area of the first page.}

\author[1]{Gordon Getty}[
                        type=editor,
                       % auid=000,bioid=1,
                       % prefix=Sir,
                        role=Researcher,
                        orcid=0000-0002-0939-6932
                        ]

% corresponding author indication
\cormark[1]
\ead{ggetty@LeakeyFoundation.org}

\author[2]{Nikita Tkachenko}[
                    type=editor,
                    % auid=000,bioid=1,
                    % prefix=Sir,
                    role=Assistant,
                    orcid=0009-0003-8681-3335
]

\ead{natkachenko@usfca.edu}

%\author[2]{Robert Trivers}[
                       % type=editor,
                       % auid=000,bioid=1,
                       % prefix=Sir,
                       % role=Researcher,
                       % orcid=0009-0003-8681-3335
%                        ]


%\ead{natkachenko@usfca.edu}


% \ead[url]{www.cvr.cc, cvr@sayahna.org}

% \credit{Conceptualization of this study, Methodology, Software}

\affiliation[1]{%
            organization={Leakey Foundation},
            addressline={1003B O’Reilly Avenue}, 
            city={San Francisco},
            postcode={94129}, 
            state={California},
            country={United States}}

\affiliation[2]{organization={Evalyn},
            addressline={2130 Fulton St}, 
            city={San Francisco},
            postcode={94117}, 
            state={California},
            country={United States}}
            
% \affiliation[2]{organization={Professor of Anthropology and Biological Sciences, Rutgers University (retired)},
           % addressline={2130 Fulton St}, 
           % city={San Francisco},
           % postcode={94117}, 
           % state={California},
           % country={United States}
%           }

% \author[2,4]{Han Theh Thanh}[style=chinese]

%\author[2,3]{CV Rajagopal}[%
%   role=Co-ordinator,
%   suffix=Jr,
%\ead{cvr3@sayahna.org}
%\ead[URL]{www.sayahna.org}

%\credit{Data curation, Writing - Original draft preparation}

% \affiliation[2]{organization={Sayahna Foundation},
%    % addressline={}, 
%    city={Jagathy},
%    % citysep={}, % Uncomment if no comma needed between city and postcode
%    postcode={695014}, 
%    state={Trivandrum},
%    country={India}}

\cortext[cor1]{Corresponding author}
% \cortext[cor2]{Principal corresponding author}

%\fntext[fn1]{This is the first author footnote. but is common to third
%  author as well.}
%\fntext[fn2]{Another author footnote, this is a very long footnote and
%  it should be a really long footnote. But this footnote is not yet
%  sufficiently long enough to make two lines of footnote text.}

%\nonumnote{This note has no numbers. In this work we demonstrate $a_b$
%  the formation Y\_1 of a new type of polariton on the interface
%  between a cuprous oxide slab and a polystyrene micro-sphere placed
%  on the slab.
%  }

\begin{abstract}
Capital growth, at large scales only, arrives with no help from net saving, and consequently with no help from consumption constraint. Net saving, at large scales, is sacrifice of consumption with nothing in return. 
\end{abstract}

% \includegraphics{figs/grabs.pdf}

%\item Research highlights item 1
%\item Research highlights item 2
%\item Research highlights item 3

% Keywords
\begin{keywords}
Consumption \sep Net saving \sep Market-value capital 
\end{keywords}

\maketitle

\section{Introduction}

Our "Sources of capital growth"\footnote{\cite{getty2024sourcescapitalgrowth}} proposed the term "thrift theory" to mean the commonly accepted doctrine that net saving, within the technological growth rate ("warranted rate"), gives capital growth. "Free growth theory" was defined as an opposite hypothesis that capital growth is explained largely or wholly by a gain in productivity of capital and labor already present, so that net saving is largely or wholly superflous. Data for net saving and market-value capital taken from national accounts of 92 nations were argued to support free growth theory. Here we pursue a parallel argument and test from data for consumption.

\section{Consumption and net saving}

Thrift theory begins with the idea that net saving \(S_{net}\), within the warranted rate, gives capital growth \(\Delta K\).\footnote{\cite{getty2024sourcescapitalgrowth} questioned the doctrine that saving equals investment, but equated them in that paper by using the word "saving" in the Keynesian sense of saving actually invested. We apply the same meaning here.} That is,
\begin{equation}
    \Delta K =S_{net} \ , \quad \text{in thrift theory.} \label{eq:1}
\end{equation}  
where "within the warranted rate" is implicit. We divide this by capital (normalize) to avoid nonsense correlation when finding averages over time. Define \(g(K) = \frac{\Delta K}{K}\) and \(s^* = \frac{S_{net}}{K}\) to infer, from Eq. \eqref{eq:1}, 
\begin{equation}
    g(K) = s^* \quad \text{in thrift theory.} \label{eq:2}
\end{equation}

In thrift theory and common teaching, where innovation provides a required condition for capital growth, but only insofar as net income is saved in investment, and where consumption and net saving sum to net income; it follows that
any change in net saving implies an equal and opposite change in consumption \(C\).\footnote{i.e., a suspension of consumption provides the only source of a gain in net saving, arguendo, while a gain in consumption becomes the only disposition of net income not applied to net saving.} That is,
\begin{equation}
    \Delta S_{net} = - \Delta C\ ,\quad \text{in thrift theory.} \quad \text{Define } c^* = \frac{C}{K} \text{ to proceed with} \label{eq:3}
\end{equation}
\vspace{-0.5cm}
\begin{equation}
    \Delta g(K) = \Delta s^* = - \Delta c^* \ , \quad \text{from which} \quad \frac{-\Delta c^*}{\Delta g(K)} = 1\ ,\quad \text{in thrift theory.}\footnotemark \label{eq:4}
\end{equation}
\footnotetext{Asterisks distinguish \(s^*\) and \(c^*\) from the Keynesian saving rate \(s\) and consumption rate \(c\), which divide by output rather than by capital.}
For convenience, define
\begin{equation}
    \theta_{c} = \frac{-\Delta c^*}{\Delta g(K)}\ . \quad \text{By Eq. \eqref{eq:4}, then,} \quad \theta_{c} = 1 \quad \text{in thrift theory.} \label{eq:5}
\end{equation}
\(\theta\) in general is what we call the "thrift index," which measures the success of the predictions of thrift theory in Eqs. \eqref{eq:4} and \eqref{eq:5} against a standard of unity (the number one), and \(\theta_{c}\) does so indirectly from data for consumption. Displays below give observations of \(\theta_{c}\) from data for \(C\) and market-value capital \(K\) in national accounts in 90 countries over the period 1980-2021.

\begin{figure}[h!]
    \centering
    \includegraphics[width=0.8\textwidth]{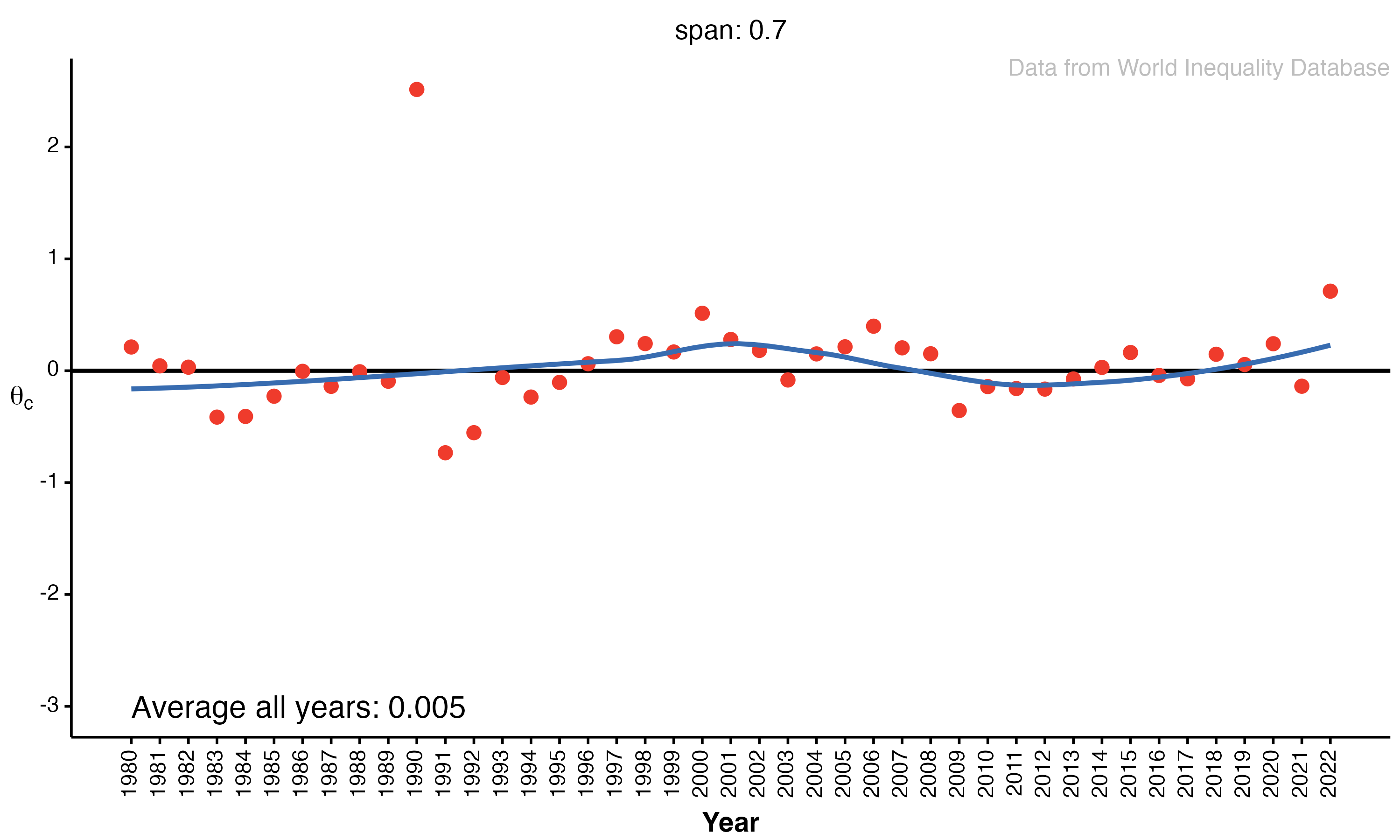}
{\raggedright%
\caption{Plots of $\theta_{c}$ averaged over all countries, GDP-weighted. Screen: \(|\Delta g(K)| \geq 0.01\). \(H_0:\ \theta_c \cong 1\).}}
\end{figure}

\begin{table}[pos=H]
    \caption{Regressions, all countries and years, GDP-weighted. \(H_0: \text{Regression} \cong 1\).}
\begin{tabular}{lcc}
\toprule
 &
$-c^*$ on $g(K)$ &
$-\Delta c^*$ on $\Delta g(K)$ \\
\midrule
Regression &
-0.181 &
-0.129 \\
Std. error &
0.048 &
0.021 \\
\midrule
Observations &
    1801 &
    772\\
$R^2$ &
    0.8 &
    0.404\\
Within $R^2$ &
    0.021 &
    0.18\\
\midrule
Screen &
    $|g(K)| \geq 0.01$ &
    $|\Delta g(K)| \geq 0.01$\\
\midrule
Year fixed effects &  Yes & Yes\\
Country fixed effects & Yes & Yes\\
\bottomrule
\end{tabular}

\end{table}

\FloatBarrier

\begin{table}[H]
    \caption{\(\theta_{c}\) in countries and periods shown. Screen: \(|\Delta g(K)| \geq 0.01\). \(H_{0} : \theta_{c} \cong 1.\)}
\centering
\begin{minipage}[t]{0.45\textwidth}
  \centering

\begin{tabular}{lrl}
\toprule
Country & $\theta_{c}$ & Periods\\
\midrule
Armenia & 0.29 & 97-19(11)\\
Aruba & 1.85 & 00-00(1)\\
Australia & -0.04 & 80-17(17)\\
Austria & -0.07 & 98-21(8)\\
Azerbaijan & -2.02 & 98-19(12)\\
\addlinespace
Bahrain & -0.24 & 10-12(2)\\
Belgium & -0.01 & 97-21(9)\\
Bolivia & -0.02 & 00-13(5)\\
Botswana & 0.19 & 97-97(1)\\
Brazil & -0.03 & 00-19(8)\\
\addlinespace
British Virgin Islands & 12.02 & 97-97(1)\\
Bulgaria & -0.25 & 98-16(8)\\
Burkina Faso & 0.04 & 01-18(8)\\
Cameroon & 1.31 & 98-19(9)\\
Canada & 0.06 & 80-20(19)\\
\addlinespace
Cape Verde & -1.47 & 10-16(5)\\
Chile & -0.14 & 00-21(11)\\
China & 0.00 & 93-19(14)\\
Colombia & 0.09 & 97-21(7)\\
Costa Rica & -0.09 & 14-19(3)\\
\addlinespace
Croatia & -0.20 & 00-21(10)\\
Curaçao & 1.72 & 03-16(7)\\
Cyprus & -0.13 & 97-21(11)\\
Czechia & -0.19 & 95-16(10)\\
Côte d’Ivoire & 1.52 & 97-00(2)\\
\addlinespace
Denmark & 0.15 & 99-21(10)\\
Dominican Republic & -0.19 & 07-15(5)\\
Ecuador & -1.17 & 10-18(6)\\
Egypt & 2.24 & 99-15(7)\\
El Salvador & 3.32 & 16-19(4)\\
\addlinespace
Estonia & -0.04 & 00-21(9)\\
Finland & -0.07 & 98-21(10)\\
France & 0.01 & 87-21(13)\\
Germany & -0.00 & 80-21(16)\\
Greece & -0.36 & 97-21(14)\\
\addlinespace
Guatemala & 0.46 & 07-21(6)\\
Guinea & -0.16 & 07-10(3)\\
Honduras & 0.27 & 05-15(5)\\
Hong Kong & 0.09 & 97-10(6)\\
Hungary & -0.04 & 97-21(12)\\
\addlinespace
Iceland & -0.09 & 02-13(8)\\
India & 0.10 & 00-15(7)\\
Indonesia & 0.02 & 19-19(1)\\
Iran & 1.21 & 99-16(9)\\
Ireland & -0.03 & 99-21(11)\\
\bottomrule
\end{tabular}

  {\footnotesize
   \raggedright
   Number of periods reported is shown in ().\par
  }

\end{minipage}
\hfill
\begin{minipage}[t]{0.45\textwidth}
  \centering

\begin{tabular}{lrl}
\toprule
Country & $\theta_{c}$ & Periods\\
\midrule
Israel & 0.58 & 00-17(7)\\
Italy & 0.14 & 86-21(11)\\
Japan & -0.10 & 85-21(13)\\
Kazakhstan & 0.02 & 97-22(13)\\
Kuwait & 0.00 & 06-15(5)\\
\addlinespace
Kyrgyzstan & 0.02 & 99-21(10)\\
Latvia & -0.09 & 00-21(9)\\
Lithuania & -0.10 & 00-21(10)\\
Luxembourg & -0.00 & 99-21(12)\\
Malaysia & -0.07 & 08-15(5)\\
\addlinespace
Malta & -0.01 & 97-21(11)\\
Mauritius & -0.06 & 17-18(2)\\
Mexico & -0.02 & 97-21(9)\\
Moldova & -0.20 & 97-19(12)\\
Mongolia & -0.08 & 11-21(4)\\
\addlinespace
Morocco & 0.49 & 00-21(8)\\
Netherlands & -0.01 & 97-21(11)\\
New Zealand & 0.03 & 97-18(11)\\
Nicaragua & 0.05 & 10-19(5)\\
Niger & 0.73 & 98-18(9)\\
\addlinespace
Norway & 0.06 & 82-19(17)\\
Peru & -0.19 & 10-21(4)\\
Philippines & 1.60 & 99-22(11)\\
Poland & 0.31 & 03-21(7)\\
Portugal & -0.05 & 97-22(10)\\
\addlinespace
Qatar & -0.21 & 04-18(6)\\
Romania & -0.17 & 98-17(7)\\
Russia & -0.09 & 99-18(6)\\
Senegal & 1.13 & 21-21(1)\\
Serbia & 0.25 & 00-21(11)\\
\addlinespace
Slovakia & -0.14 & 01-21(9)\\
Slovenia & -0.10 & 99-21(11)\\
South Africa & 0.16 & 99-21(9)\\
South Korea & 0.07 & 99-19(8)\\
Spain & 0.07 & 97-21(12)\\
\addlinespace
Sweden & -0.03 & 81-21(22)\\
Switzerland & -0.03 & 93-20(14)\\
Tunisia & 3.39 & 97-06(4)\\
Turkey & 0.03 & 13-15(2)\\
USA & -0.04 & 83-20(18)\\
\addlinespace
Ukraine & 0.05 & 97-19(14)\\
United Kingdom & -0.20 & 81-21(19)\\
Uzbekistan & -0.17 & 13-21(4)\\
Vanuatu & -0.06 & 03-06(2)\\
Venezuela & -0.02 & 00-12(6)\\
\bottomrule
\end{tabular}
\end{minipage}
\end{table}

\FloatBarrier

\section{Assessment of test results}

Average \(\theta_{c}\) was found at 0.005, and regressions of \(-c^*\) on \(g(K)\) and \(-\Delta c^*\) on \(\Delta g(K)\) were found at -0.181 and -0.129 respectively. The value expected in thrift theory for each of these measurements is 1.0 before allowance for market noise. These test results, we think, justify the interpretation that net saving through consumption restraint contributes essentially nothing to capital growth.

\cite{getty2024sourcescapitalgrowth} comparing data for capital growth to data for net saving rather than consumption, found \(\theta\) at 0.064, and found positive regressions at 0.059 and 0.077. We suggested that these findings could have other explanations than those given by thrift theory. Our findings in this paper again suggest other causes.

\section{Balanced growth}

\cite{keynes1936} taught that households tend to consume a constant proportion of income. Consumption would tend to peak somewhere between a worker's mid-forties and retirement, then, as income does.
Franco Modigliani, in 1954,\footnote{\cite{modigliani_brumberg_1954}} proposed conversely
that younger households tend to save to afford children and retirement, and that older ones dissave to pay consumption costs of both as they arrive. 
Milton Friedman suggested in 1957 that households consider lifetime income, not current income alone, and choose between saving and consumption accordingly.

In any of these three scenarios, random distribution of ages could allow phases of more saving in some households to coincide with phases of more consumption in other households. It becomes possible, then, if net saving realizes equal capital growth, for consumption, net saving, capital growth and capital itself all to grow at the same rate for all households together. Then net output \(Y_{net}\) could grow at that same rate if the capital/output ratio holds constant. This scenario where \(C\), \(S_{net}\), \(K\) and \(Y_{net}\) all grow at the same rate is called balanced growth.

We agree that households save more and consume more in phases, that net saving tends to bring equal capital growth at the household scale (see Section 5 below), and that random distribution of phases could explain balanced growth within thrift theory. The observation that the consumption/capital ratio \(c^*\) does seem to hold more or less constant over scale and time could therefore fit the predictions of thrift theory, at first glance, as well as of free growth theory.

A closer look tells the difference. First, perfectly balanced growth is mathematically possible only where capital growth rate \(g(K)\) holds constant.\footnote{Because net output/income gives the only source of capital growth, and consumption gives the only avenue of capital depletion after depreciation is recaptured within the value-added chain.} That would imply \(\Delta g(K) = 0\). The record shows no period where \(\Delta g(K)\) equals zero or \(c^*\) equals unity exactly. Even when absolute \(\Delta g(K)\) is small, and \(c^*\) is near unity, Eqs. \eqref{eq:4} and \eqref{eq:5} still apply in thrift theory. Here too, \(\theta_c\) is observed to approach zero rather than unity.

Second, for reasons shown in Section 4, we screen out years where absolute \(\Delta g(K)\) is less than 0.01. Thus our tests show only results where the predictions of thrift theory and free growth theory differ substantially. 

Test results are conclusive. \(\theta_{c}\) and the regressions of  \(-c^*\) on \(g(K)\) and of \(-\Delta c^*\) on \(\Delta g(K)\) are all predicted at unity in thrift theory. They are measured respectively at near zero and substantially less than zero. These findings parallel and reinforce our similar ones from data for net saving and consumption reported in \cite{getty2024sourcescapitalgrowth}.

%
% \begin{table}
% \centering
% \caption{Yearly \(\theta_c\) and \(g(K)\) in Countries shown\footnotemark } \label{tbl-3}
% \begin{tabular}{rcc|cc|cc|cc|cc|cc|cc|rcc|cc|cc|cc|cc|cc|cc|rcc|cc|cc|cc|cc|cc|cc|rcc|cc|cc|cc|cc|cc|cc|rcc|cc|cc|cc|cc|cc|cc|rcc|cc|cc|cc|cc|cc|cc|rcc|cc|cc|cc|cc|cc|cc|rcc|cc|cc|cc|cc|cc|cc|rcc|cc|cc|cc|cc|cc|cc|rcc|cc|cc|cc|cc|cc|cc|rcc|cc|cc|cc|cc|cc|cc|rcc|cc|cc|cc|cc|cc|cc|rcc|cc|cc|cc|cc|cc|cc|rcc|cc|cc|cc|cc|cc|cc|rcc|cc|cc|cc|cc|cc|cc|}
% \toprule
% \multicolumn{1}{c}{ } & \multicolumn{2}{c}{U.K.} & \multicolumn{2}{c}{U.S.} & \multicolumn{2}{c}{Germany} & \multicolumn{2}{c}{France} & \multicolumn{2}{c}{Italy} & \multicolumn{2}{c}{Spain} & \multicolumn{2}{c}{Japan} \\
% \cmidrule(l{3pt}r{3pt}){2-3} \cmidrule(l{3pt}r{3pt}){4-5} \cmidrule(l{3pt}r{3pt}){6-7} \cmidrule(l{3pt}r{3pt}){8-9} \cmidrule(l{3pt}r{3pt}){10-11} \cmidrule(l{3pt}r{3pt}){12-13} \cmidrule(l{3pt}r{3pt}){14-15}
%  & $\theta_{c}$ & $g(K)$ & $\theta_{c}$ & $g(K)$ & $\theta_{c}$ & $g(K)$ & $\theta_{c}$ & $g(K)$ & $\theta_{c}$ & $g(K)$ & $\theta_{c}$ & $g(K)$ & $\theta_{c}$ & $g(K)$\\
% \input{tables/tbl-3}
% \end{tabular}
% \end{table}
% \footnotetext{Values for other countries show in our web appendix \href{https://web-appendix.shinyapps.io/Sources\_of\_Capital\_Growth/}{web appendix (https://web-appendix.shinyapps.io/Sources\_of\_Capital\_Growth)}.}
% \footnotetext{Unweighted average absolute \(\theta_c\) and \(g(K)\) overall years.}
%

\section{Screens against small denominators}

No measurement is perfect. Small mismeasurements of small denominators, even if numerators were measured accurately, can distort calculations of fractions dramatically. The denominator in calculating \(\theta_c\) is \(\Delta g(K)\), the change in capital growth rate. This change, up or down, is occasionally very small. We control for that risk by screening out years where absolute \(\Delta g(K)\) shows as less than \(0.01\).

\section{Growth by thrift at smaller scales}

Free growth theory predicts only at the collective or global scale, meaning the scale of all households and nations taken together. We follow John \cite{rae1834} in reasoning that some households and communities can grow by saving if others dissave as much. We accept that much growth at the scale of individual households and firms, and possibly even small nations, fits the predictions of thrift theory. 

\section{Source of data}

All our data are drawn from \href{https://wid.world/document/distributional-national-accounts-guidelines-2020-concepts-and-methods-used-in-the-world-inequality-database/}{Distributional National Accounts (DINA)} shown in the free online database \href{wid.world.com}{World
Inequality Database (WID)}. This source collates data from national accounts and tax data
of 105 countries in constant currency units, and adjusts them where needed to conform to current standards of the System of National Accounts
(SNA) published by the United Nations. In this paper, we show results for the 90 of
those countries which report 
all information needed for deriving \(\theta_{c}\). We find market-value capital \(K\) from WID as Capital Wealth (mnweal). We construct consumption as the sum of the consumption expenditures of the government (mcongo) and private expenditures of households and NPISH (mconhn).
GDP, which we use only for weighting purposes, is reproduced from GDP (mgdpro).

\section{Summary and discussion}

Tests of thrift theory from data for consumption support findings from data for net saving in \cite{getty2024sourcescapitalgrowth}. This paper refers to that one for logical arguments and policy prescriptions. The finding is that capital growth at very large scales, not small ones, is financed from depreciation investment with no help from net saving. Thus net saving, we conclude, is sacrifice of consumption with nothing in return. 

These findings warrant attention. Much of macroeconomics as taught today, whether in Keynesian or anti-Keynesian schools, rests on the assumption that capital growth requires \textit{equal} net saving. Data show that it requires \textit{no} net saving. We suggest urgent review of our data and methods, as well as new tests from other data, and study of implications if our findings are confirmed.

% \begin{figure}[h]
%     \centering
%     \includegraphics[width=1\textwidth]{plots/theta1.png}
%     \caption{Plots of $\theta_{c1}$}
% \end{figure}

% \begin{figure}[h]
%     \centering
%     \includegraphics[width=1\textwidth]{plots/theta3.png}
%     \caption{Plots of $\theta_{c3}$}
% \end{figure}

% \begin{figure}[h]
%     \centering
%     \includegraphics[width=1\textwidth]{plots/theta4.png}
%     \caption{Plots of $\theta_{c4}$}
% \end{figure}

\printcredits

\bibliographystyle{cas-model2-names}

\bibliography{cas-refs}

\end{document}